CORRESPONDENCE

To the Editors of 'The Observatory'

*The sunspot observations by Rheita in 1642*

The Maunder Minimum (MM) was a period of very low solar activity that occurred from 1645 to 1715 approximately[1]. It is the only grand minimum of solar activity that has been observed in the telescope era and, therefore, scientists are very interested in that episode of the history of our Sun[2-3]. Hoyt and Schatten[4] provided a database of observed sunspot groups that is the only daily index to study solar activity around the MM. Some studies have tried to improve this database. For example, Vaquero et al.[5] added, changed, and removed some records for the period 1637-1642, just before the onset of the MM. In particular, they changed the interpretation of the sunspot observations recorded by A. M. Rheita. According to Hoyt and Schatten[4], Rheita observed eight sunspot groups from 9 to 21 February 1642. However, Vaquero et al.[5] stated that Rheita observed one sunspot group during 9-22 June 1642. Zolotova and Ponyavin[6] have argued recently that "in February 1642, Rheita reported eight sunspot groups, but all other observers registered a fewer number of groups. Thus, Cycle -10 can be high or in the middle." The correct interpretation of this fragment of Rheita has crucial importance in estimating the amplitude of this solar cycle just before the MM. Therefore, we show here that this record has been misinterpreted by Zolotova and Ponyavin[6], presenting the original Latin text and a modern English translation.

The original text, located on pages 242-243 of the book entitled *Oculus Enoch et Eliae*[7], and the modern English translation are the following:

*Certè quod iam diximus, propria experientiâ, Coloniae, anno 1642, experti sumus, dum ingentem stellarum solarium turmam maiorum et minorum, per 14 dies et ultra, sibi inuicem continuâ serie succedentium, cum stupore, solarem discum adeò occupare vidimus, ut lux eius, maximè media et intensissima, haud leuiter illis fuerit hebetata. Nam tubo optimo, in medio Solaris, disci globum perfectissimè rotundum, subnigrum,*

*pugni magnitudinem quasi excedentem conspeximus, idque directissimo aspectu; qui et per octiduum Solis haud exiguam portionem eclipsauit, maximasque aëri turbationes, utpotè ventos, imbres et frigora, in medio iunii attulit. Prout crebris observationibus iam à multis annis compertum habemus scilicet ferè semper aëris insigniores et magis notabiles mutationes ex dictarum stellarum solarem discum subeuntium agmine contingere et euenire.*

Certainly, what we have just said [that sunlight is weakened by the appearance of sunspots] was checked by our own experience in Cologne in 1642, when, during 14 days and more we saw with astonishment that a lot of larger and smaller solar starts, passing over the sun one after another in continuous succession, occupied so much the solar disk light, especially its central light and intense, was not a little attenuated by them. Indeed, with a telescope perfectly suited we contemplate amidst the solar disk a perfectly round, blackish circle, almost exceeding the size of a fist, and this with total clarity; and this circle for eight days covered a not small part of the sun and caused major disruptions in weather: wind, rain, and cold in mid-June. Under frequent observations for many years, we have full confidence that almost always the most marked and significant weather changes occur due to the conjunction of the solar planets when superimposed above the solar disk.

From the original text, it is clear that (*i*) there are not eight sunspot groups on the solar disc but eight days when one large sunspot covered a not small part the sun; (*ii*) Rheita observed in mid-June instead of mid-February; and (*iii*) Rheita did not provide an exact number of sunspot groups on the solar disc (maybe he was referring to one large group or maybe a chain of two or three consecutive sunspot groups).

We appreciate the support of Junta de Extremadura and Ministry of Economy and Competitiveness (AYA2011-25945).

Yours faithfully,

JUAN M. GÓMEZ

Departamento de Ciencias de la Antigüedad

Facultad de Filosofía y Letras, Universidad de Extremadura

Avda. de la Universidad s/n

10003 Cáceres, Spain

*JOSÉ M. VAQUERO*

Departamento de Física

Centro Universitario de Mérida, Universidad de Extremadura

Avda. Santa Teresa de Jornet, 38

E-06800 Mérida, Badajoz, Spain

2015 February 24